\documentclass[aps,prl,nofootinbib,twocolumn,superscriptaddress]{revtex4-1}
\usepackage[pdftex]{hyperref}
\usepackage{euscript}
\usepackage{amssymb}
\usepackage{amsfonts}
\usepackage{amsbsy}
\usepackage{epsfig}
\usepackage{subfigure}
\usepackage{amsthm}
\usepackage{amscd}
\usepackage{amstext}
\usepackage{verbatim}
\usepackage{amsmath}
\usepackage{url}
\usepackage{bm}
\usepackage{color}

\begin{document}

\title{Deep learning black hole metrics from shear viscosity}
\author{Yu-Kun Yan}
\email{ykyan.phys@gmail.com}
\affiliation{Department of physics, Shanghai University, Shanghai, 200444,
China}
\affiliation{School of Physics, University of Chinese Academy of Sciences, Beijing,
100049, China}

\author{Shao-Feng Wu}
\email{sfwu@shu.edu.cn}

\affiliation{Department of physics, Shanghai University, Shanghai, 200444,
China}
\affiliation{Center for Gravitation and Cosmology, Yangzhou University, Yangzhou 225009,
China}

\author{Xian-Hui Ge}
\email{gexh@shu.edu.cn}

\affiliation{Department of physics, Shanghai University, Shanghai, 200444,
China}
\affiliation{Center for Gravitation and Cosmology, Yangzhou University, Yangzhou 225009,
China}

\author{Yu Tian}
\email{ytian@ucas.ac.cn}

\affiliation{School of Physics, University of Chinese Academy of Sciences, Beijing,
100049, China}
\affiliation{Institute of Theoretical Physics, Chinese Academy of Sciences, Beijing
100190, China}
\affiliation{Center for Theoretical Physics, Massachusetts Institute of Technology, MA
02139, Cambridge, USA}
\affiliation{Center for Gravitation and Cosmology, Yangzhou University, Yangzhou 225009,
China}

\begin{abstract}
Based on AdS/CFT correspondence, we build a deep neural network to
learn black hole metrics from the complex frequency-dependent shear
viscosity. The network architecture provides a discretized representation of
the holographic renormalization group flow of the shear viscosity and can be applied to a large class of strongly coupled field theories. Given the
existence of the horizon and guided by the smoothness of spacetime, we show
that Schwarzschild and Reissner-Nordstr\"{o}m metrics can be learned
accurately. Moreover, we illustrate that the generalization ability of the
deep neural network can be excellent, which indicates that by using the black
hole spacetime as a hidden data structure, a wide spectrum of the shear
viscosity can be generated from a narrow frequency range. These results are
further generalized to an Einstein-Maxwell-dilaton black hole. Our work
might not only suggest a data-driven way to study holographic transports but
also shed some light on holographic duality and deep learning.
\end{abstract}

\maketitle

\emph{Introduction}.---Renormalization group (RG) is a physical scheme to
understand various emergent phenomena in the world through iterative coarse
graining \cite{Kadanoff1966,Wilson74,Wilson83,Polchinski84}. Deep learning
(DL) is the core algorithm of the recent wave of artificial intelligence
\cite{Sejnowski2018}. It has been suggested that RG and DL might have a
logic in common \cite{Wen16,Mehta1401} and their relation has attracted a
lot of interest \cite%
{Beny13,Saremi13,Paul1412,Braddea1610,Lin1608,Ringel1704,Oprisa1705,Foreman1710,Iso1801,Wang1802}%
.

RG is believed to be one of the key elements to understand quantum gravity. In
particular, by anti-de Sitter/conformal field theory (AdS/CFT)
correspondence \cite{Maldacena97,Gubser98,Witten98,Susskind98}, a strongly
coupled quantum critical theory in the d-dimensional spacetime is
reorganized along the RG scale, inducing a classical theory of gravity in
the (d+1)-dimensional AdS spacetime. RG is not the only connection between DL
and gravity. Through the study of tensor networks \cite%
{Bridgeman1603,Scholl2011,Vidal2011}, especially the multi-scale
entanglement renormalization ansatz (MERA) \cite{Vidal0512}, it has been
realized that the way the geometry is emergent from field theories usually
involves the network and optimization, which are two important ingredients
of DL.

Because of these connections, the deep neural network (DNN) may be capable of
providing a research platform for holographic duality \cite%
{Shu1705,You1709,Hashimoto1802,Hashimoto1809,You1903,Hashimoto1903,Hartnoll1906,Tan1908}%
. One can expect at least two benefits: It is helpful to understand how the
spacetime emerges and can be used to build a data-driven phenomenological model for
strongly coupled field theories. The latter was initiated in \cite%
{Hashimoto1802}, where the inverse problem of the AdS/CFT is studied, that
is, how to reconstruct the spacetime metric from the given field theory data
by the DNN which implements the AdS/CFT. Subsequently, the so-called AdS/DL
correspondence is applied to learn the bulk metric from the lattice QCD data
of the finite-temperature chiral condensate. Interestingly, the emergent
metric exhibits both a black hole horizon and an IR wall with finite height,
signaling the crossover of QCD thermal phases \cite{Hashimoto1809}.

Let us briefly introduce the prototype of AdS/DL (i.e., the first
numerical experiment in \cite{Hashimoto1802}). It establishes the
architecture of the DNN according to the discretized equation of motion of
the $\phi ^{4}$ theory minimally coupled to gravity. The training data
are the one-point function and the conjugate source with a label determined
by the near-horizon scalar field. The target of learning is the metric of
the Schwarzschild black hole. A key technique is to design the
regularization by which the emergent metric is favored to be smooth. It is
found that the DNN performs better near the boundary than near the horizon
where the relative error is around 30\%. In \cite{Tan1908}, it has been
attempted to learn the Reissner-Nordstr\"{o}m (RN) metric by AdS/DL but the
mean square error (MSE) ranges from $\mathcal{O}\left( 10^{-3}\right) $ to $%
\mathcal{O}\left( 10^{-1}\right) $. Importantly, it was revealed that the
form of the regularization term must be fine-tuned\ for different metrics.
This suggests that their DNN may not find the target metric
if it is unknown previously, since it is difficult to\ judge which is closer
to the target metric under different regularizations.

In this paper, we will extend the physical range of the AdS/DL nontrivially
and illustrate that it can be realized without previous technical problems.
Our strategies are as follows. First, AdS/CFT is almost customized for the
computation of the transports of strongly coupled quantum critical systems
at finite temperatures \cite{ZaanenBook}. In particular, the application of
holography is anchored partially in the prediction of the nearly perfect
fluidity \cite{KSS}, which has been observed in the hot quark gluon plasmas
and cold unitary Fermi gases \cite{Schaefer0904}. With these in mind, we
adapt the complex frequency-dependent shear viscosity as the given field
theory data. Second, we propose to build the DNN according to the
holographic RG flow of the shear viscosity. Up to the holographic
renormalization, this flow was presented in the well-known holographic
membrane paradigm \cite{Liu0809,Strominger1006}, which interpolates the
standard AdS/CFT correspondence and the classical black hole membrane
paradigm smoothly \cite{Thorne86,Wilczek97}. Third, we assume the existence
of the horizon, which will reduce the learning difficulties. Fourth, the
system error in \cite{Hashimoto1802} comes from adding labels on the data
and introducing the regularization. Because the horizon value of the shear
response is completely determined by the regularity analysis on the horizon,
we can generate the data by the flow from IR to UV. The direction of
information transfer is contrary to \cite{Hashimoto1802,Hashimoto1809} and
there is no error caused by the labels of the data. Fifth, we still use the
regularization to guide the network to find a smooth metric. However, our
training process has two stages and the regularization is only required in
the first stage, so we can choose any regularization term as long as it
leads to a smaller loss in the second stage. Finally, we will discuss
possible extensions and physical implications.

\emph{From RG flow to DNN.}---Suppose that a strongly coupled field theory
is dual to the (3+1)-dimensional Einstein gravity minimally coupled with
matter, which admits a homogeneous and isotropic (along the field theory
directions) black hole solution with the metric ansatz%
\begin{equation}
ds^{2}=-g_{tt}(r)dt^{2}+g_{rr}(r)dr^{2}+g_{xx}(r)d\vec{x}^{2}.
\label{BH ansatz}
\end{equation}%
When the black hole is perturbed by time-dependent sources, the shear mode $%
\left( \delta g\right) _{\;x_{2}}^{x_{1}}=h(r)e^{-i\omega t}$ of the
gravitational wave is controlled by the equation of motion%
\begin{equation}
\frac{1}{\sqrt{-g}}\partial _{r}(\sqrt{-g}g^{rr}\partial _{r}h)+g^{tt}\omega
^{2}h=0,
\end{equation}%
provided that the graviton is massless\footnote{%
The gauged coordinate-invariance symmetries in the bulk demand that the
gravitons are massless and they are dual to the global spacetime symmetries
on the boundary \cite{ZaanenBook}.} \cite{Hartnoll1601}. In the Hamiltonian
form, the wave equation can be written as%
\begin{eqnarray}
\Pi  &=&-\sqrt{-g}g^{rr}\partial _{r}h,  \label{PI1} \\
\partial _{r}\Pi  &=&\sqrt{-g}g^{rr}g^{tt}\omega ^{2}h,  \label{PI2}
\end{eqnarray}%
where $\Pi $ is the momentum conjugate to the field $h$. Consider the
foliation in the $r$-direction and define the shear response function $\chi
=\Pi /(i\omega h)$ on each cutoff surface. Substituting Eq. (\ref{PI1}) into
Eq. (\ref{PI2}), one can obtain%
\begin{equation}
\partial _{r}\chi -i\omega \sqrt{\frac{g_{rr}}{g_{tt}}}\left( \frac{\chi ^{2}%
}{g_{xx}}-g_{xx}\right) =0.  \label{floweq1}
\end{equation}%
Note that this radial flow equation has been derived in \cite{Liu0809} where
the DC limit is focused on\footnote{%
In the Wilsonian formulation, the flow equation can be retrieved as the $%
\beta $-functions of double-trace couplings \cite%
{Son1009,Polchinski1010,Liu1010}.}. We will study the frequency-dependent
behavior.

Applying the regularity of $\chi $ on the horizon, one can read off the
horizon value of $\chi $ directly%
\begin{equation}
\chi (r_{h})=g_{xx}(r_{h}),  \label{chiIR}
\end{equation}%
where $r_{h}$ is the horizon radius. Taking Eq. (\ref{chiIR}) as the IR
boundary condition, the flow equation can be integrated to the UV boundary. However,
it should be pointed out that the response function $\chi $ on the UV\
boundary is not equal to the shear viscosity $\eta $ of the boundary field
theory. In the Supplementary Material (SM), we will clarify the {relation}
between them using the Kubo formula of the complex frequency-dependent shear
viscosity \cite{Read1207,Wu2015} and the holographic renormalization \cite%
{Henningson9806,HJ}. It can be found that for a large class of holographic
models, including the Einstein-Maxwell theory which will be studied below,
the {relation} is%
\begin{equation}
\eta (\omega )=\left. \chi (\omega ,r)+i\omega r\right\vert _{r\rightarrow
\infty }.  \label{etachi}
\end{equation}

In the metric ansatz (\ref{BH ansatz}), $g_{xx}$ can be fixed as $r^{2}$
without loss of generality but $g_{tt}$ and $g_{rr}$ are independent in
general. However, there are some black holes which share the feature $%
g_{tt}g_{rr}=1$, indicating that the radial pressure is the negative of the
energy density \cite{Jacobson0707}. For simplicity, we will consider this
situation first and return to the more general case later. Thus, the metric
ansatz can be reduced to%
\begin{equation}
ds^{2}=\frac{1}{z^{2}}\left[ -f(z)dt^{2}+\frac{1}{f(z)}dz^{2}+d\vec{x}^{2}%
\right] ,  \label{fmetric}
\end{equation}%
where we have used the coordinate $z=r_{h}/r$ so that the horizon is located at $z=1$ and
the boundary at $z=0$. Accordingly, Eq. (\ref{floweq1}) can be rewritten as%
\begin{equation}
\left( \eta -\frac{i\omega }{z}\right) ^{\prime }+\frac{i\omega }{f}\left[
z^{2}\left( \eta -\frac{i\omega }{z}\right) ^{2}-\frac{1}{z^{2}}\right] =0,
\label{floweq2}
\end{equation}%
where we have set $r_{h}=1$ and the prime denotes the derivative with
respect to $z$. Note that we have replaced $\chi (\omega ,z)$ with $\eta
(\omega ,z)-i\omega /z$ from IR to UV. Compared to Eq. (\ref{etachi}) where
the replacement occurs only on the UV, we have found that this technique
reduces the discretized error considerably. The radially varying function $%
\eta (\omega ,z)$ can be referred to the holographic RG flow of the shear
viscosity. In the following, we will build a DNN according to the flow
equation (\ref{floweq2}).

\begin{figure}[tbp]
\centering
\includegraphics[width=8cm]{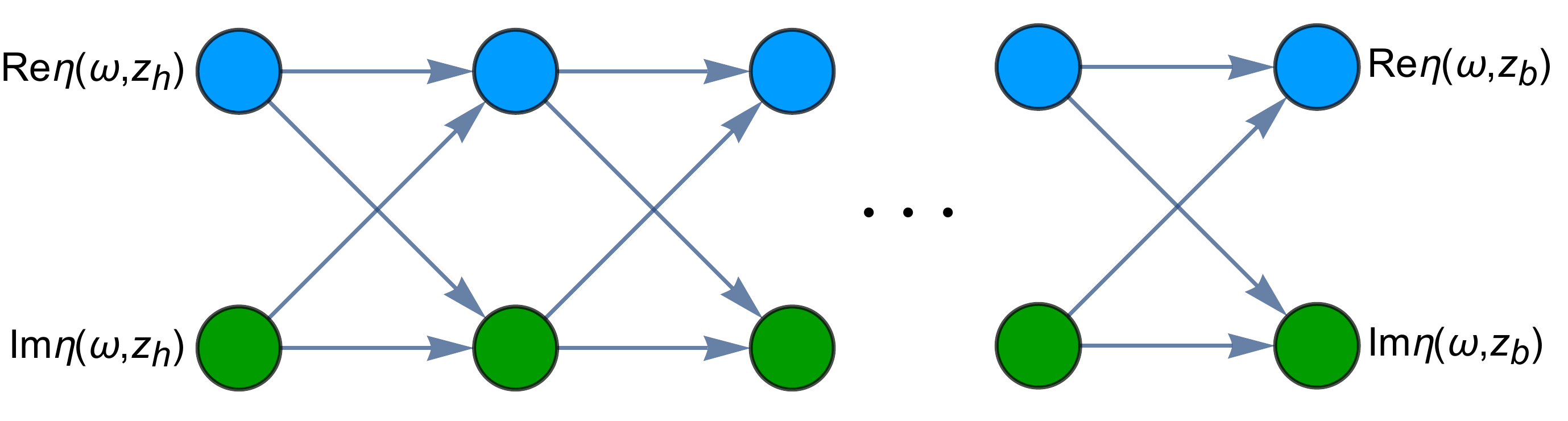}
\caption{The architecture of the DNN. The green and blue nodes have $N$
layers, which propagate the shear viscosity from IR to UV by discretized RG
flow equations (\protect\ref{Dflow}). The arrows indicate the direction of
information transfer.}
\label{fig1}
\end{figure}

FIG. \ref{fig1} is a schematic diagram of the DNN. The $N$ deep layers are
located by discretizing the radial direction%
\begin{equation}
z(n)=z_{b}+n\Delta z,\;\Delta z=\frac{z_{h}-z_{b}}{N},
\end{equation}%
where $z_{b}$ ($z_{h}$) is the UV (IR) cutoff and the integer $n $ belongs
to $\left[ 1,N\right] $. The trainable weights of the DNN represent the
discretized metrics. The input is $\eta (\omega ,z_{h})$ and the output is $%
\eta (\omega ,z_{b})$. The information is transferred from the $N$th layer
(IR) to the $1$th layer (UV). The propagation rule of $\eta (\omega ,z)$
between layers is determined by the discretized representation of Eq. (\ref%
{floweq2}):%
\begin{eqnarray}
\text{\textrm{Re}}\eta \left( z+\Delta z\right) &=&\text{\textrm{Re}}\eta
\left( z\right) \left[ 1+\Delta z\frac{2\omega z^{2}}{f\left( z\right) }%
\left( \text{\textrm{Im}}\eta \left( z\right) -\frac{\omega }{z}\right) %
\right] ,  \notag \\
\text{\textrm{Im}}\eta \left( z+\Delta z\right) &=&\text{\textrm{Im}}\eta
\left( z\right) +\Delta z\frac{\omega z^{2}}{f\left( z\right) }\Big[\frac{%
1-f(z)}{z^{4}}  \notag \\
&&-\left( \text{\textrm{Re}}\eta \left( z\right) \right) ^{2}+\left( \text{%
\textrm{Im}}\eta \left( z\right) -\frac{\omega }{z}\right) ^{2}\Big].
\label{Dflow}
\end{eqnarray}%
Here we have separated the discretized flow equation into real and imaginary
parts for the convenience in DL.

The loss function we choose is the $L^{2}$-norm%
\begin{equation}
L_{\mathrm{DNN}}=\sum_{\mathrm{data}}\left\vert \eta (\omega ,z_{b})-\bar{%
\eta}(\omega ,z_{b})\right\vert ^{2},  \label{loss}
\end{equation}%
up to a regularization term, if existent. Here $\eta $ represents the input
data and $\bar{\eta}$ is what the DNN generates.

We need a regularization term which can guide the DNN to find a smooth
black hole metric. In principle, the form of the regularization term can be
arbitrary as long as it can reduce the final loss. In practice, our
regularization term\ is specified as%
\begin{eqnarray}
L_{\text{\textrm{REG}}} &=&c_{1}\sum_{n=1}^{N-1}\frac{1}{z(n)^{c_{2}}}\left[
f(z(n+1))-f(z(n))\right] ^{2}  \notag \\
&&+c_{3}\left[ f(z(N))-0\right] ^{2},  \label{reg}
\end{eqnarray}%
where the two parts are designed for the smoothness of the metric and the
existence of the horizon, respectively. Three hyperparameters $c_{1},c_{2}$,
and $c_{3}$ are introduced.
\begin{figure}[tbp]
\centering
\subfigure[]{
  \includegraphics[width=4cm]{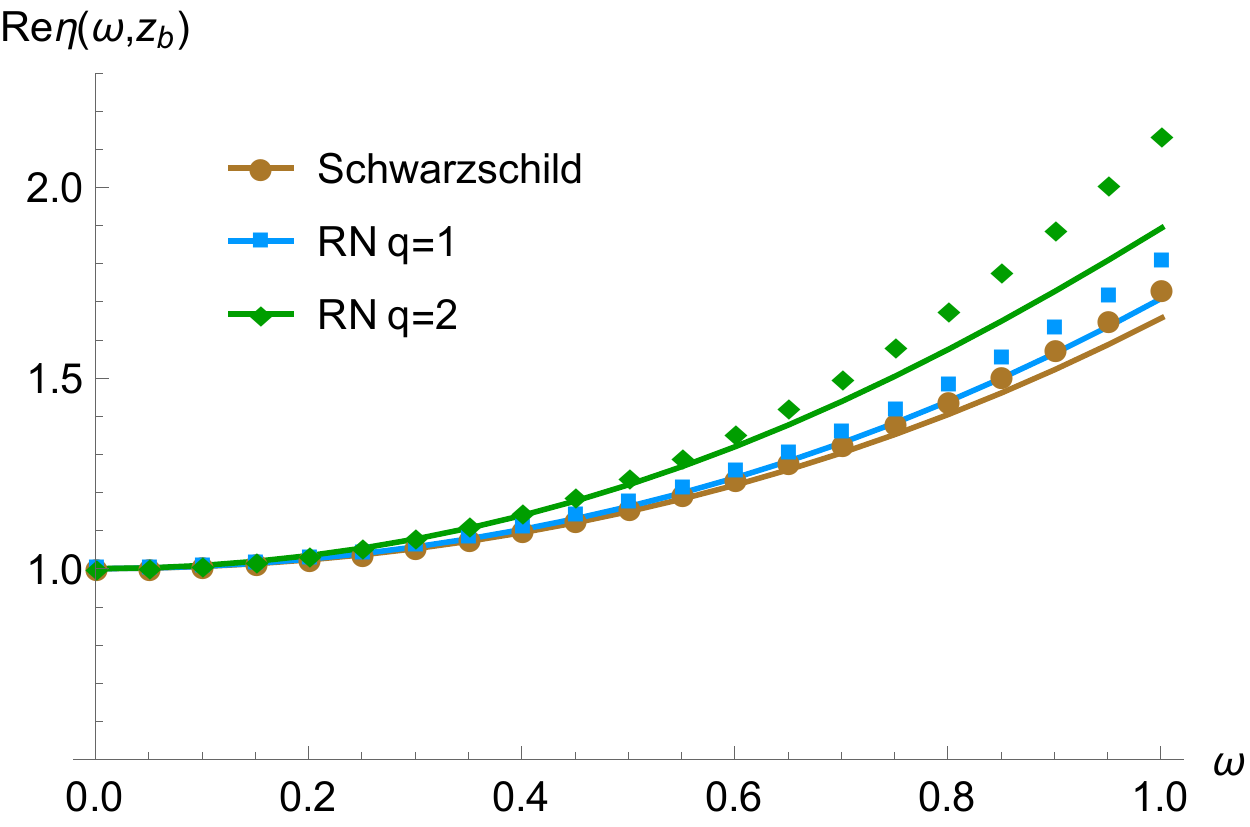}}
\subfigure[]{
  \includegraphics[width=4cm]{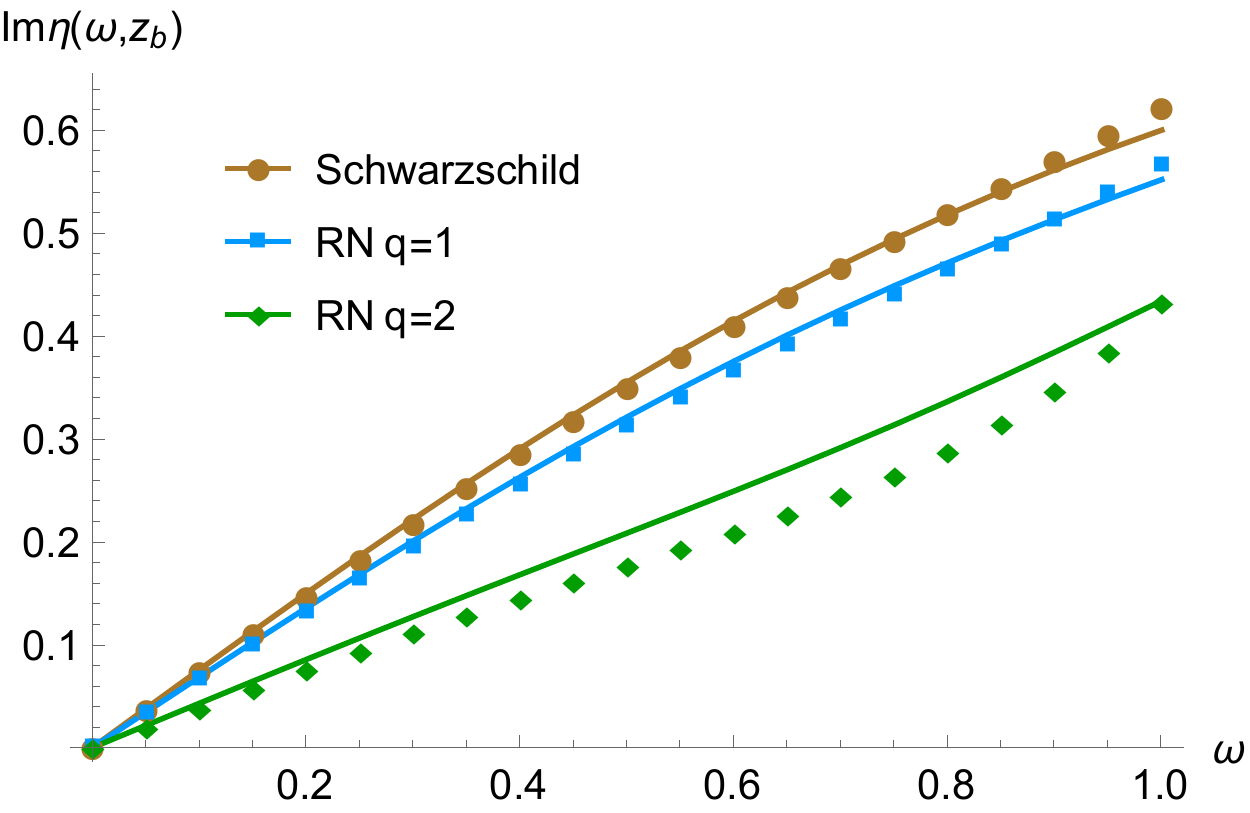}}
\caption{The data of the shear viscosity generated by different metrics. (a)
and (b) denote the real and imaginary parts, respectively. The curves are
generated by the continuous holographic RG flow equation of the shear
viscosity, while the markers are generated by the discretized equation.}
\label{fig2}
\end{figure}

\emph{Generated data and discretized error.}---We specify the discretized RG
flow and hence the DNN by setting $z_{b}=0.01$, $z_{h}=0.99$, and $N=10$.
Using the discretized flow equation (\ref{Dflow}) with the IR boundary
condition $\eta (\omega ,z_{h})=1+i\omega $ and the target metric, we can
generate the data required by DL. Here we consider the two most famous black
holes, i.e. the Schwarzschild and RN. They are characterized respectively by
the functions%
\begin{eqnarray}
f(z) &=&1-z^{3},  \label{SWC metric} \\
f(z) &=&1-z^{3}-\frac{q^{2}z^{3}}{4}+\frac{q^{2}z^{4}}{4},  \label{RN metric}
\end{eqnarray}%
where $q$ is the charge density. We generate 2000 data $(\omega ,\ \eta
(\omega ,z_{b}))$ with even frequency spacing. The training set and
validation set account for 90\% and 10\%, respectively. In Fig. (\ref{fig2}%
), we compare the data with the ones generated by numerically solving the
continuous flow equation of the shear viscosity. It is found that the
discretized error is small when the target is the Schwarzschild
metric, and increases with the frequency and the charge density. The
discretized error can be reduced by adding more layers but it requires
powerful computing capabilities. As a proof of principle, here we simply
assume that the discretized error does not affect our results qualitatively%
\footnote{%
Recently, the discretized error has been taken into account for the
application of AdS/DL to the QCD experimental data \cite{Hashimoto2005}.}.

\begin{figure}[tbp]
\centering
\subfigure[]{
  \includegraphics[width=4cm]{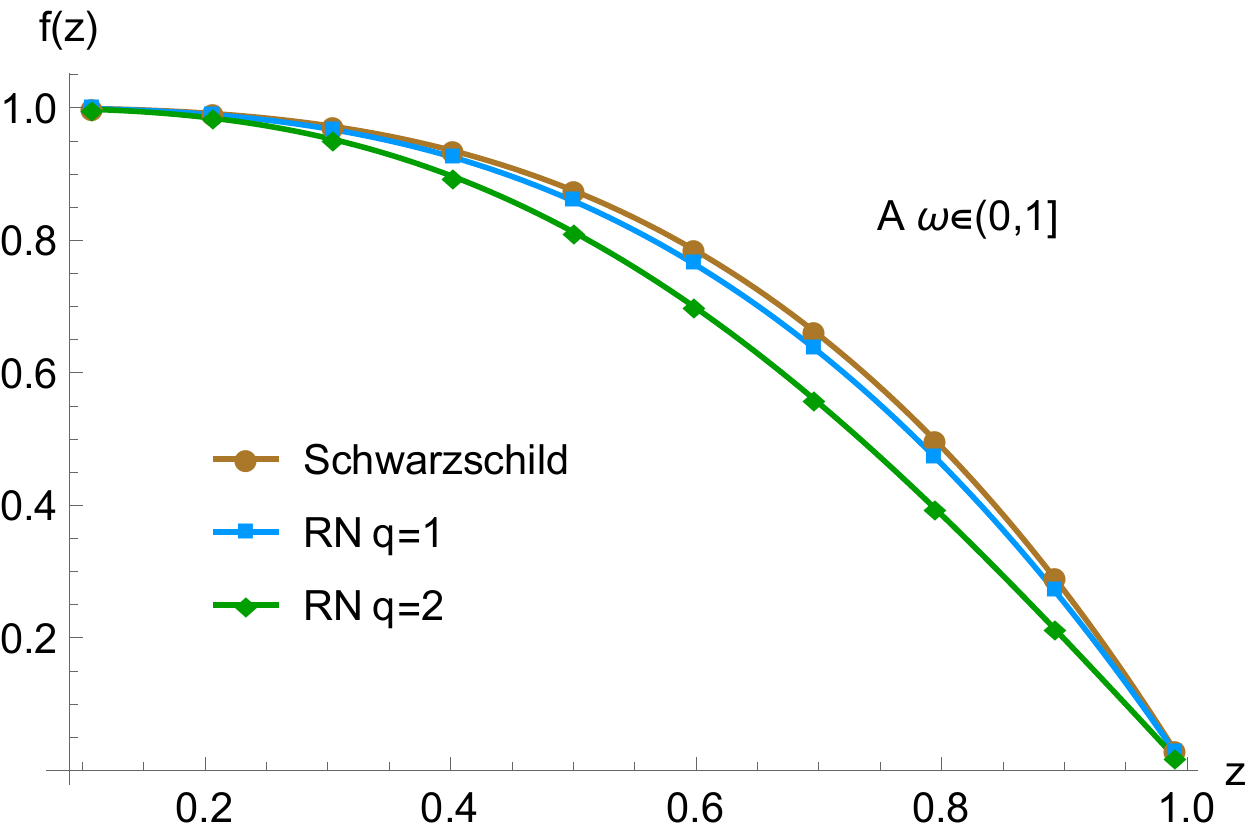}}
\subfigure[]{
  \includegraphics[width=4cm]{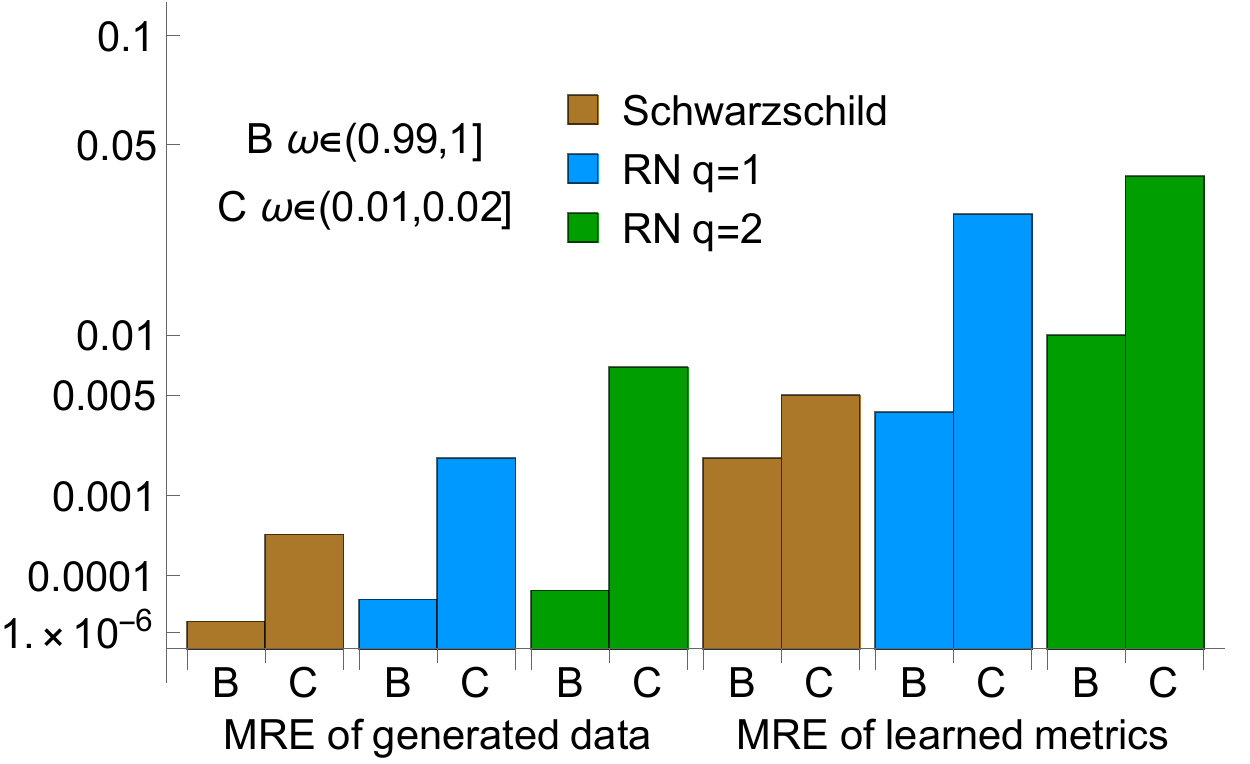}}
\caption{The performance of the DNN. (a) The curves are the target metrics
and the markers are the results learned from the data A, which has a wide
frequency range. (b) The right half of the bars represents the MRE of the
metrics which are learned from the data with narrow ranges B and C. The left
half represents the MRE of the wideband data, which are generated using the
metrics learned from the narrowband.}
\label{fig3}
\end{figure}

\emph{Emergence and generalization.}---With the data in hand, we will train
the DNN and extract the weights. The training scheme will be given in the
SM. In TABLE S.1, we list the training reports after two training stages of
various numerical experiments. Among others, it is shown that from the
dataset with $\omega \in (0,1]$\footnote{%
The frequency is measured in units of the horizon radius.}, the
Schwarzschild and RN metrics can be learned with high accuracy: the mean
relative error (MRE) is around $0.1\%$. Note that the MSE is $\mathcal{O}%
\left( 10^{-7}\right) $. The target and learned metrics have been plotted in
FIG. (\ref{fig3}.a).

Hereto, we almost naively select the frequency range of the data as $\Delta
\omega =1$.\ One important question in DL is how well the model generalizes.
To proceed, we consider different datasets with the narrow frequency range $%
\Delta \omega =10^{-2}$ and keep each of them with 2000 data. Interestingly,
we find that both Schwarzschild and RN metrics still can be well learned,
although the error will increase when the frequency window is close to zero
and especially when the charge density is large. This is shown by the MRE of
the metrics learned from two typical windows, see the right half of FIG. (%
\ref{fig3}.b). Furthermore, it suggests that the generalization ability of
the DNN can be excellent. Indeed, in the left half of FIG. (\ref{fig3}.b),
we illustrate that using the metric learned from the data with $\Delta
\omega =10^{-2}$, one can generate the data with $\Delta \omega =1$ very
accurately. In the best performing example, the MRE of the generated data
can be $\mathcal{O}\left( 10^{-6}\right) $. We also note that the examples
with relatively large errors in FIG. (\ref{fig3}.b) can be expected because
the DC limit of the shear viscosity is determined solely by the physics on
the horizon. In particular, when the charge density increases, the RN black
hole approaches extremality and the IR CFT associated with the \textrm{AdS}$%
_{2}\times \mathrm{R}^{2}$ geometry gradually begins to dominate the
low-frequency physics \cite{Liu0907,Edalati0910}. Similarly, we do not
expect that the DNN can learn well from a very high-frequency window, where
the UV\ CFT associated with the AdS boundary should dominate\footnote{%
In fact, it has been observed in \cite{Matteo1903,Matteo1910} that the
retarded Green function for the shear stress operator at the infinite
frequency is determined by the energy density. We thank Matteo Baggioli for
the discussion on this point.}.

\emph{Beyond Einstein-Maxwell.}---Our DNN is not only applicable to the
Einstein-Maxwell theory. In the SM, we demonstrate in general that the DNN
can be applied to the Einstein-Maxwell-dilaton (EMD) theory with the typical
potential and coupling of the dilaton \cite{Gubser0911}. The only nontrivial
constraint is that the conformal dimension of the dilaton operator $\Delta
_{\phi }$ should be less than 5/2. To be more specific, we further study a
concrete EMD theory. It admits the analytical black hole solution with $%
\Delta _{\phi }=2$ and its holographic renormalization has been given in
\cite{Kim1608}. This exemplifies our general argument explicitly. We also
carry out the numerical experiments as before. It is found that the
performance of the DNN for the EMD black hole is similar to that for the RN,
see TABLE S.1. Note that at the zero-temperature limit, the EMD black hole
becomes a special hyperscaling violating Lifshitz geometry with the
asymptotic AdS. Moreover, here the metric components $g_{tt}(r)g_{rr}(r)\neq
1$. So what the DNN has learned is the joint factor%
\begin{equation}
f(z)=\left. \frac{1}{r^{2}}\sqrt{\frac{g_{tt}(r)}{g_{rr}(r)}}\right\vert
_{r=r_{h}/z},
\end{equation}%
as we will explain below.

\emph{Viscosity and entanglement.}---A more general metric like the EMD
black hole has two independent components $g_{tt}$ and $g_{rr}$. From Eq. (%
\ref{floweq1}), one can find that they appear in the form of the joint
factor $g_{rr}/g_{tt}$. Therefore, without the constraint on the
energy-momentum tensor for $g_{tt}g_{rr}=1$, the DNN still can be applied to
learn the joint factor, but in general each of them cannot be learned
separately from the shear response. Nevertheless, if there is another way to
determine one component, the other can be obtained by the DNN. For example,
there is evidence that the entanglement plays an important role in weaving
the spacetime \cite%
{Maldacena0106,RT0603,Swingle0905,Raamsdonk1010,Susskind1306}. Among others,
it has been shown that the holographic entanglement entropy $S(l)$ can be
used to fix the bulk metric wherever the extremal surface reaches \cite%
{Bilson0807}, which can be described as%
\begin{equation}
\left. r\sqrt{g_{rr}(r)}\right\vert _{r=r_{h}/z}=\frac{1}{2\pi ^{2}L}%
z^{2}\partial _{z}\int_{z_{b}}^{z}\frac{z_{\ast }S(z_{\ast })}{\sqrt{%
z^{4}-z_{\ast }^{4}}}dz_{\ast },
\end{equation}%
where we have set the gravitational constant $16\pi G=1$ and $z_{\ast }$ is
determined by $S^{\prime }(l)=8\pi L/z_{\ast }^{2}$. Note that $l$ and $L$
denote the finite width and the (regularized) infinite length of the
rectangle on which the extremal surface is anchored \cite{RT0603}. Since the
holographic entanglement entropy is only related to $g_{rr}$, it can
complement to the shear viscosity to determine two metric components.

\emph{Conclusion and discussion.}---Using a simple DL algorithm, we studied
an inverse problem of AdS/CFT: Given the complex frequency-dependent shear
viscosity of boundary field theories at finite temperatures, can the bulk
metrics of black holes be extracted? We showed that Schwarzschild, RN, and
EMD metrics can be learned by the DNN with high accuracy. The network
architecture can be taken as a discretized representation of the holographic
RG flow of the shear viscosity, hence supporting the underlying {relationship%
} among DL, RG, and gravity. We pointed out that our DNN is applicable to any
strongly coupled field theory provided that: it is dual to the
(3+1)-dimensional Einstein gravity minimally coupled with matter, it allows a
homogeneous and isotropic black hole solution, the graviton mass in the wave
equation vanishes, and the UV relation (\ref{etachi}) holds. The extensions
to the symmetry-breaking situations, the higher spacetime dimensions, and
the modified theories of gravity should be worthwhile. Among others, we note
that using the wave equation with the graviton mass which has been built up
in \cite{Hartnoll1601}, one can construct the RG flow and the DNN where the
graviton mass is encoded into new trainable weights. It is interesting to
study whether the DNN can learn the metric and the mass simultaneously. In
addition to various extensions, there are two open questions which should be
mentioned. (i) Is there a better ansatz for the regularization term? Note
that the regularization in this work is not to prevent overfitting as usual
in machine learning. Instead, it is a guide to the minimum loss. We might
need a deeper physical understanding of the regularization. (ii) How to
reduce the discretized error at high frequencies and low temperatures
sufficiently? Compared with directly increasing the number of layers, a more
efficient method might be to apply the recently proposed DNN models of
ordinary differential equations \cite{Chen1806}. We ultimately hope that our
work could suggest a data-driven way to study holographic transports.

Moreover, we found that the complete black hole metric from IR to UV can be
well learned from the data with narrow frequency ranges. We also checked
that the performance of the DNN is hardly changed by randomly deleting
several data points in our numerical experiments. These two facts indicate
that the shear viscosity encodes the spacetime in a very different way from
the entanglement entropy. The latter probes the deeper spacetime only by the
$S(l)$ with a larger $l$, so any data point is necessary to reconstruct the
spacetime. Perhaps we can describe the difference concisely as follows: the
non-local observable (entanglement entropy) on the boundary probes the bulk
spacetime locally, while the local observable (shear viscosity) probes it
non-locally.

\begin{figure}[tbp]
\centering
\includegraphics[width=8cm]{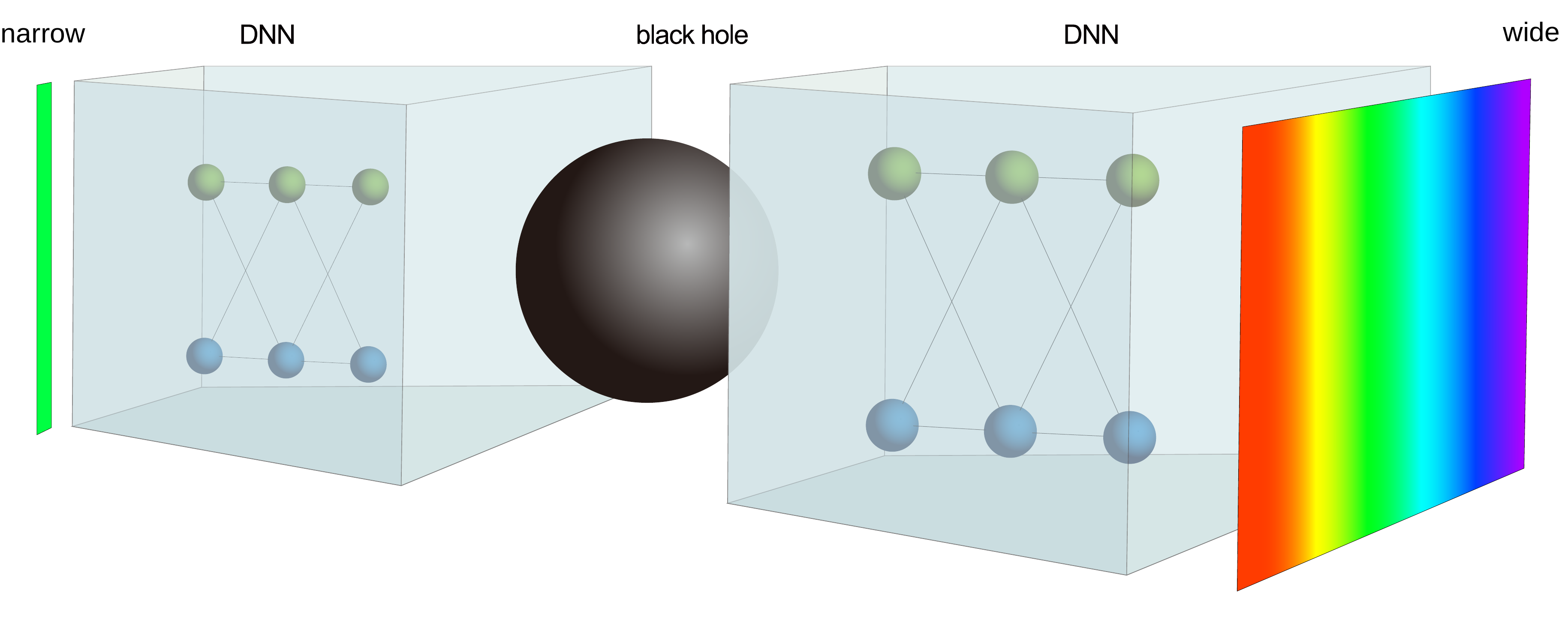}
\caption{Schematic diagram: the DNN encodes the black hole spacetime, by
which a wide spectrum of the field theory data can be generated from its
narrow piece.}
\label{fig4}
\end{figure}

Furthermore, this non-locality leads to the excellent generalization ability
of the DNN, which should be important in the application to the experimental
data collected only in a part of the spectrum. Theoretically, from the
perspective of machine learning, it usually implies that the data are highly
structured\footnote{%
Another possibility is that the network has some symmetry \cite{Zhai1901}.}.
This structure is often important but obscure\footnote{%
For example, using the generative adversarial network (GAN), the approximate
statistical predictions have been made recently in the string theory
landscape \cite{Halverson2001}, where the accurate extrapolation capability
has been exhibited on simulating K\"{a}hler metrics. It was speculated that
this is the first evidence of Reid's fantasy: all Calabi-Yau manifolds with
fixed dimension are connected.}, due to the infamous black-box problem of
machine learning. However, here the structure is nothing but the
higher-dimensional black hole spacetime. This strong emergence might shed
some light on the understanding of DL. Last but not least, the excellent
generalization suggests that the strongly coupled field theories with
gravity dual could exhibit another feature of the hologram\ in addition to
encoding the higher dimension: The local (a small piece of the hologram) can
reproduce the whole, see the schematic diagram FIG. (\ref{fig4}).

\emph{Acknowledgments.}---We thank Koji Hashimoto for reading the draft and
giving valuable comments. We also thank Matteo Baggioli, Yongcheng Ding,
Wei-Jia Li, Tomi Ohtsuki, and Fu-Wen Shu for helpful discussions. SFW and
XHG are supported by NSFC grants No. 11675097 and No. 11875184,
respectively. YT is supported partially by NSFC grants No. 11975235 and No.
11675015. He is also supported by the \textquotedblleft Strategic Priority
Research Program of the Chinese Academy of Sciences\textquotedblright\ with
Grant No. XDB23030000.


\onecolumngrid
\newpage

\setcounter{equation}{0} \setcounter{table}{0}\setcounter{figure}{0}%
\setcounter{page}{1}\renewcommand{\theequation}{S.\arabic{equation}} %
\renewcommand{\thetable}{S.\arabic{table}} \makeatletter

\begin{center}
\textbf{\large Supplementary material for\\[0pt]
`Deep learning black hole metrics from shear viscosity'}
\end{center}

\section{Complex frequency-dependent shear viscosity}

Compared to the real shear viscosity at the zero frequency limit, the study
of the complex frequency-dependent counterpart is rare. So let's begin from
reviewing the Kubo formula of the complex frequency-dependent shear
viscosity. In \cite{Read1207}, the Kubo formulas for the stress-stress
response function at zero wavevector is derived from first principles. The
approach given in \cite{Read1207} starts from a microscopic Hamiltonian and
define the viscosity tensor as the linear response to a uniform external
strain. In \cite{Wu2015}, an alternative field-theory approach is proposed,
by which the Ward identity of viscosity coefficients in \cite{Read1207} is
retrieved and extended. Here we will follow \cite{Wu2015} to give the
definition of the complex frequency-dependent shear viscosity by the
generating functional.

For a theorist, the viscosity can be measured by sending a gravitational
wave through the system \cite{Son2007}. Suppose that a homogeneous and
isotropic system lives in the two-dimensional flat space which is perturbed
by a uniform gravitational wave. The response tensor $Y^{ijkl}$ can be
defined by%
\begin{equation}
\delta \langle T^{ij}(t)\rangle _{\mathrm{r}}=-\frac{1}{2}\int dt^{\prime
}Y^{ijkl}(t-t^{\prime })\partial _{t^{\prime }}\delta g_{\mathrm{r}%
kl}(t^{\prime }).  \label{TX}
\end{equation}%
Here the subscript $\mathrm{r}$ and the subscript $\mathrm{a}$ below
indicate that we have invoked the closed time-path formalism to discuss the
real-time response. The elastic modulus and the viscosity tensor can be
further defined by separating the right hand of Eq. (\ref{TX}) into two
parts,%
\begin{equation}
\delta \langle T^{ij}(t)\rangle _{\mathrm{r}}=-\frac{1}{2}\int dt^{\prime
}\lambda ^{ijkl}(t-t^{\prime })\delta g_{\mathrm{r}kl}(t^{\prime })-\frac{1}{%
2}\int dt^{\prime }\eta ^{ijkl}(t-t^{\prime })\partial _{t^{\prime }}\delta
g_{\mathrm{r}kl}(t^{\prime }).
\end{equation}%
The stress tensor can be derived by the variation of the generating
functional with respect to the metric%
\begin{equation}
\langle T^{ij}(t)\rangle _{\mathrm{r}}=\frac{2}{\sqrt{g}}\frac{\delta W}{%
\delta g_{\mathrm{a}ij}(t)},
\end{equation}%
and the second variation leads to the retarded correlator%
\begin{equation}
G_{\mathrm{ra}}^{ij,kl}(t)\equiv \frac{4\delta W}{\delta g_{\mathrm{a}%
ij}(t)\delta g_{\mathrm{r}kl}(0)}=\delta ^{kl}\langle T^{ij}\rangle \delta
(t)-\lambda ^{ijkl}(t)-\partial _{t}\eta ^{ijkl}(t).  \label{GLY}
\end{equation}%
The elastic modulus is the stress response up to the zeroth-order in time
derivatives, which can be determined by the constitute relation of perfect
fluids. In hydrodynamic expansion, one has%
\begin{equation}
\delta \langle T^{ij}(t)\rangle _{\mathrm{r}}=-\left( P\delta ^{ik}\delta
^{jl}+\frac{1}{2}\delta ^{ij}\delta ^{kl}\kappa ^{-1}\right) \delta g_{%
\mathrm{r}kl}(t),  \label{hydro1}
\end{equation}%
where $P$ is the pressure and $\kappa ^{-1}$ is the inverse compressibility.
Then the elastic modulus can be given by
\begin{equation}
\lambda ^{ijkl}(t)=\left[ P\left( \delta ^{ik}\delta ^{jl}+\delta
^{il}\delta ^{jk}\right) +\delta ^{ij}\delta ^{kl}\kappa ^{-1}\right] \delta
(t).  \label{lambda1}
\end{equation}%
The viscosity tensor can be decomposed as%
\begin{equation}
\eta ^{ijkl}(t)=\zeta (t)\delta ^{ij}\delta ^{kl}+\eta (t)\left( \delta
^{ik}\delta ^{jl}+\delta ^{il}\delta ^{jk}-\delta ^{ij}\delta ^{kl}\right)
+\eta ^{H}(t)\left( \delta ^{jk}\epsilon ^{il}-\delta ^{il}\epsilon
^{kj}\right) .  \label{etaijkl}
\end{equation}%
The coefficients $\zeta ,\eta ,\eta ^{H}$ denote the bulk, shear, and Hall
viscosities, respectively. Substituting the last two equations into Eq. (\ref%
{GLY}), one can obtain%
\begin{eqnarray}
G_{\mathrm{ra}}^{12,12}(t) &=&-\lambda ^{1212}(t)-\partial _{t}\eta
^{1212}(t)  \notag \\
&=&-P\delta (t)-\partial _{t}\eta (t).
\end{eqnarray}%
In Fourier space, we have the Kubo formula of the shear viscosity%
\begin{equation}
\eta (\omega )=\frac{G_{\mathrm{ra}}^{12,12}(\omega )+P}{i\omega }.
\label{Kubo}
\end{equation}%
Note that this formula applies to the complex shear viscosity at all
frequencies. In contrast, many literatures focus on the real part and the DC
limit of the shear viscosity, so the pressure in Eq. (\ref{Kubo}) is often
neglected.

\section{Holographic renormalization}

We proceed to bridge the shear viscosity $\eta (\omega )$ to the shear
response $\chi (\omega )$. The essential procedure is to carry out the
holographic renormalization \cite{Henningson9806}. Consider that the bulk
action includes the Einstein gravity and the minimally coupled matter%
\begin{equation}
S_{\mathrm{bulk}}=\int d^{4}x\sqrt{-g}\left( R+6+L_{\mathrm{matter}}\right) ,
\label{bulkEMA}
\end{equation}%
where we have set the AdS radius $L=1$ and the Newton constant $16\pi
G_{N}=1 $. Suppose that the background metric is homogeneous and isotropic
along the field theory directions%
\begin{equation}
ds^{2}=-g_{tt}(r)dt^{2}+g_{rr}(r)dr^{2}+g_{xx}(r)d\vec{x}^{2}.
\end{equation}%
Accordingly, the energy-momentum tensor can be written by%
\begin{equation}
T_{\mu \nu }=\mathrm{diag}\left(
T_{tt}(r),T_{rr}(r),T_{xx}(r),T_{xx}(r)\right) .
\end{equation}%
Perturbing the Einstein equation on the background, the wave equation of the
shear mode is derived as \cite{Hartnoll1601}%
\begin{equation}
\frac{1}{\sqrt{-g}}\partial _{r}(\sqrt{-g}g^{rr}\partial _{r}h)+g^{tt}\omega
^{2}h=0,  \label{wave eq}
\end{equation}%
where we have assumed that the square of the graviton mass $%
m^{2}=g^{xx}T_{xx}-\delta T_{xy}/\delta g_{xy}$ is vanishing. Without loss
of generality, we will set $g_{xx}=r^{2}$ hereafter. We further require that
the metric can be expanded near the AdS boundary, with the form%
\begin{eqnarray}
g_{tt} &=&r^{2}(1+\frac{a_{1}}{r}+\frac{a_{2}}{r^{2}}+\frac{a_{3}}{r^{3}}%
+\cdots ),  \notag \\
g^{rr} &=&r^{2}(1+\frac{b_{1}}{r}+\frac{b_{2}}{r^{2}}+\frac{b_{3}}{r^{3}}%
+\cdots ),  \label{metric2}
\end{eqnarray}%
where $a_{i}$ and $b_{i}$ are some constants.

Near the boundary, the wave equation\ of the shear mode has the asymptotic
solution%
\begin{equation}
h=h^{(0)}+\frac{1}{r^{2}}h^{(2)}+\frac{1}{r^{3}}h^{(3)}+\cdots .
\label{solu2}
\end{equation}%
Here $h^{(0)}$ is the source, $h^{(2)}$ is fixed by $h^{(0)}$ as $%
h^{(2)}=h^{(0)}\omega ^{2}/2$, and $h^{(3)}$ relies on $h^{(0)}$ and the
incoming boundary condition on the horizon. In solving the asymptotic
equation, one can find%
\begin{equation}
\left( a_{1}+b_{1}\right) \omega ^{2}h^{(0)}=0.
\end{equation}%
The situation $a_{1}=-b_{1}\neq 0$ is rare, if existed. So we focus on $%
a_{1}=-b_{1}=0$.

We write down the Gibbons-Hawking term and the counterterms%
\begin{eqnarray}
S_{\mathrm{GH}} &=&-2\int d^{3}x\sqrt{-\gamma }K,  \label{GHEMA} \\
S_{\mathrm{ct}} &=&\int d^{3}x\sqrt{-\gamma }\left( -4+R+L_{\mathrm{matter}%
}^{(1)}\right) ,  \label{ctEMA}
\end{eqnarray}%
where $\gamma ^{ab}$ is the induced metric, $K$ is the external curvature,
and $L_{\mathrm{matter}}^{(1)}$ is contributed by the matter.

We expand the on-shell bulk action, the Gibbons-Hawking term and the
counterterms to the quadratic order of the shear mode,%
\begin{eqnarray}
&&\left. S_{\mathrm{bulk}}+S_{\mathrm{GH}}+S_{\mathrm{ct}}\right\vert _{%
\mathrm{on-shell,\;quadratic}}  \notag \\
&=&\int d^{2}x\int_{-\infty }^{\infty }\frac{d\omega }{2\pi }\frac{1}{2}%
\left[ \left( -\frac{r^{2}\omega ^{2}}{\sqrt{g_{tt}}}+4r^{2}\sqrt{g_{tt}}-2r%
\sqrt{\frac{g_{tt}}{g_{rr}}}-\frac{r^{2}g_{tt}^{\prime }}{\sqrt{g_{tt}g_{rr}}%
}+L_{\mathrm{matter}}^{(2)}\right) \bar{h}h-r^{2}\sqrt{\frac{g_{tt}}{g_{rr}}}%
\bar{h}h^{\prime }\right] ,  \label{qaction}
\end{eqnarray}%
where $\bar{h}$ has the argument $-\omega $ and $L_{\mathrm{matter}}^{(2)}$
denotes the matter contribution which may be divergent on the boundary.

Substituting the asymptotic solution (\ref{solu2}) and the metric (\ref%
{metric2}) into Eq. (\ref{qaction}), we obtain the renormalized action:%
\begin{equation}
S_{\mathrm{ren}}=\int d^{2}x\int_{-\infty }^{\infty }\frac{d\omega }{2\pi }%
\frac{1}{2}\left[ (3a_{3}-2b_{3}+L_{\mathrm{matter}}^{(3)})\bar{h}%
^{(0)}h^{(0)}+3\bar{h}^{(0)}h^{(3)}\right] ,
\end{equation}%
where $L_{\mathrm{matter}}^{(3)}$ is contributed by the matter and it is
finite.

Invoking the holographic dictionary, one can extract the retarded correlator
from $S_{\mathrm{ren}}$:%
\begin{equation}
G_{\mathrm{ra}}^{12,12}(\omega )=(3a_{3}-2b_{3}+L_{\mathrm{matter}}^{(3)})+3%
\frac{h^{(3)}}{h^{(0)}}.
\end{equation}%
Expand the response function near the boundary, which yields%
\begin{equation}
\chi \equiv \frac{\Pi }{i\omega h}=-\frac{\sqrt{-g}g^{rr}\partial _{r}h}{%
i\omega h}=\left. \frac{3}{i\omega }\frac{h^{(3)}}{h^{(0)}}-i\omega
r\right\vert _{r\rightarrow \infty }.
\end{equation}%
Then we have%
\begin{equation}
G_{\mathrm{ra}}^{12,12}(\omega )=(3a_{3}-2b_{3}+L_{\mathrm{matter}%
}^{(3)})+i\omega \left( \chi +i\omega r\right) _{r\rightarrow \infty }.
\end{equation}%
Reading the pressure $P=-G_{\mathrm{ra}}^{12,12}(0)$ from Eq. (\ref{Kubo})
and the DC response $\chi (0)=r_{h}^{2}$ from Eq. (5) of the main text, we
find%
\begin{equation}
G_{\mathrm{ra}}^{12,12}(\omega )=-P+i\omega \left( \chi +i\omega r\right)
_{r\rightarrow \infty }+L_{\mathrm{matter}}^{(3)}\left( \omega \right) -L_{%
\mathrm{matter}}^{(3)}\left( 0\right) .  \label{GM}
\end{equation}%
Notice that given the Dirichlet boundary conditions as usual, $L_{\mathrm{%
matter}}^{(1)}$ should be an intrinsic scalar on the 2+1-dimensional
boundary, which indicates that we can parameterize
\begin{equation}
L_{\mathrm{matter}}^{(3)}\left( \omega \right) -L_{\mathrm{matter}%
}^{(3)}\left( 0\right) =\omega ^{2}M,  \label{LM}
\end{equation}%
where $M$ represents a finite real number.

Combining Eq. (\ref{GM}), Eq. (\ref{LM}) and Eq. (\ref{Kubo}), we obtain the
relation between shear response and shear viscosity
\begin{equation}
\eta (\omega )=\left. \chi (\omega )+i\omega r\right\vert _{r\rightarrow
\infty }-i\omega M.  \label{etachim}
\end{equation}

Some remarks on the parameter $M$ are in order. First of all, for the
Einstein-Maxwell theory, $M$ is equal to zero. Second, it is also vanishing
for some other matter fields. We take the massive scalar field $\phi $ as an
example, which is dominated by%
\begin{equation}
L_{\mathrm{matter}}=-\frac{1}{2}g^{\mu \nu }\partial _{\mu }\phi \partial
_{\nu }\phi -\frac{1}{2}m_{\phi }^{2}\phi ^{2}.
\end{equation}%
It is dual to the relevant operator with the conformal dimension $\Delta
_{\phi }=\frac{3}{2}+\sqrt{\frac{9}{4}+m_{\phi }^{2}}$ when $\Delta _{\phi
}<3$. The scalar field yields two counterterms related to $\omega ^{2}$,
that is%
\begin{equation}
L_{\mathrm{matter}}^{(1)}=\phi \nabla ^{2}\phi +R\phi ^{2},
\end{equation}%
where we have neglected two prefactors. When $\Delta _{\phi }<5/2$, these
two terms do not contribute to $M$. Third, for the bottom-up holographic
model, usually only the IR behavior of the metric\ is concerned. Instead,
the holographic renormalization depends on the UV alone. Thus, one can
assume that the target IR metric is embedded into a suitable UV background,
by which $M=0$. In the main text, we focus on the UV-complete metrics with $%
M=0$ for simplicity. More generally, one can follow this philosophy. Thus,
for a large class of strongly coupled theories with gravity dual, we have
obtained%
\begin{equation}
\eta (\omega )=\left. \chi (\omega )+i\omega r\right\vert _{r\rightarrow
\infty }.
\end{equation}

\section{Einstein-Maxwell-dilaton theory}

We will illustrate that our DNN can be applied to the
Einstein-Maxwell-dilaton (EMD) theory \cite{Gubser0911}.
Consider the bulk action in a $d+1$-dimensional spacetime%
\begin{equation}
S_{\mathrm{bulk}}=\int d^{d+1}x\sqrt{-g}\left[ R-\frac{Z\left( \phi \right)
}{4}F^{\mu \nu }F_{\mu \nu }-\frac{1}{2}\nabla _{\mu }\phi \nabla ^{\mu
}\phi -V\left( \phi \right) \right] .
\end{equation}%
Here $V\left( \phi \right) $ is the potential of the dilaton and $Z\left(
\phi \right) $ is the coupling between the dilaton and the gauge field. For
any dilaton functions, the wave equation of the shear mode is same as Eq. (%
\ref{wave eq}). The dilaton functions should be constrained to accommodate a
spacetime solution with asymptotically AdS. Typically, they can be expanded
near the AdS boundary%
\begin{eqnarray}
Z\left( \phi \right) &=&1+\cdots ,  \notag \\
V\left( \phi \right) &=&-d(d-1)+\frac{1}{2}V_{2}\phi ^{2}+\cdots ,
\end{eqnarray}%
where the ellipsis denotes the higher order terms of $\phi $. Using the gauge%
\begin{equation}
ds^{2}=d\rho ^{2}+\gamma _{ij}(\rho ,x)dx^{i}dx^{j},\;A_{\rho }=0,
\end{equation}%
we further require the boundary conditions of the fields as \cite%
{Papadimitriou0505}%
\begin{equation}
\gamma _{ij}(\rho ,x)\simeq e^{2\rho }\bar{\gamma}_{ij}(x),\;A_{i}(\rho
,x)\simeq A_{i}(x),\;\phi (\rho ,x)\simeq e^{-(d-\Delta _{\phi })\rho }\bar{%
\phi}(x),
\end{equation}%
where the conformal dimension is%
\begin{equation}
\Delta _{\phi }=\frac{d}{2}+\sqrt{\frac{d^{2}}{4}+m_{\phi }^{2}}
\label{conf dim}
\end{equation}%
with the mass square%
\begin{equation}
m_{\phi }^{2}=\left. \frac{\partial ^{2}}{\partial \phi ^{2}}\left[ V\left(
\phi \right) +\frac{Z\left( \phi \right) }{4}F^{2}\right] \right\vert _{\phi
=0}.  \label{mVZ}
\end{equation}%
According to the holographic renormalization, especially the Hamilton-Jacobi
approach \cite{HJ}, one can build up the most general ansatz for the
intrinsic counterterms on the boundary\footnote{%
Here we select the grand canonical ensemble and impose the Dirichlet
boundary condition on the dilaton as usual. If the Neumann or mixed boundary
condition is imposed, one should involve an additional boundary term which
is not intrinsic on the boundary. However, it does not contribute to the
parameter $M$, as we will show below.}%
\begin{equation}
S_{\mathrm{ct}}=-2\int d^{d}x\sqrt{-\gamma }U(\gamma ^{ij},A_{i},\phi ).
\label{U}
\end{equation}%
It can be organized in an expansion%
\begin{equation}
U=U_{(0)}+U_{(2)}+\cdots +U_{(2\left\lfloor \frac{d}{2}\right\rfloor )},
\end{equation}%
where $U_{(2k)}$ contains $k$ inverse metrics and $\left\lfloor
d/2\right\rfloor $ denotes the integer no more than $d/2$. Hereafter, we
will focus on $d=3$ and $U=U_{(0)}+U_{(2)}$.

Keeping in mind the U(1) symmetry, one can find that the contribution of
gauge fields starts from $U_{(4)}$. The dilaton may contribute to both $%
U_{(0)}$ and $U_{(2)}$. The leading terms of $U_{(2)}$ with the dilaton are $%
\phi \nabla ^{2}\phi $ and $R\phi ^{2}$. Thus, as analyzed in the previous
section, we can set $\Delta _{\phi }<5/2$ to impose the vanishing of the
matter parameter $M$.

We proceed to study a concrete EMD theory \cite{Gubser0911}. We specify the
potential and the coupling as%
\begin{equation}
Z\left( \phi \right) =\exp (\phi /\sqrt{3}),\;V\left( \phi \right) =-6\cosh
(\phi /\sqrt{3}).  \label{ZV}
\end{equation}%
This theory allows an analytical black hole solution%
\begin{eqnarray}
ds^{2} &=&u^{2}f_{2}(u)\left( -f_{1}(u)dt^{2}+dx^{2}+dy^{2}\right) +\frac{1}{%
u^{2}f_{1}(u)f_{2}(u)}du^{2},  \label{EMADmetric} \\
f_{1}(u) &=&1-\frac{1}{\left( Q+u\right) ^{3}}\left( m_{+}+Q^{3}\right) ,
\notag \\
f_{2}(u) &=&\left( 1+\frac{Q}{u}\right) ^{\frac{3}{2}},  \notag
\end{eqnarray}%
associated with the profile of matter fields%
\begin{eqnarray}
A &=&\frac{\sqrt{3Q\left( m_{+}+Q^{3}\right) }}{Q+u_{+}}\frac{u-u_{+}}{Q+u}%
dt, \\
\phi (u) &=&\frac{\sqrt{3}}{2}\log \left( 1+\frac{Q}{u}\right) ,
\end{eqnarray}%
where $m_{+}$ and $Q$ are two independent parameters. Note that $m_{+}$ is
related to the horizon radius $u_{+}$. Interestingly, when the black hole
solution approaches extremality, it interpolates the asymptotic AdS on the UV
and the conformal-to-AdS$_{2}$ geometry on the IR\ with the Lifshitz and
hyperscaling violating exponents $z=\infty \ $and $-\theta /z=1$.

Substituting the background solution into Eq. (\ref{conf dim}) and Eq. (\ref%
{mVZ}), we obtain the conformal dimension $\Delta _{\phi }=2$. Since it is
less than $5/2$, we can infer the matter parameter $M=0$. Moreover, the
holographic renormalization of this theory has been studied in \cite{Kim1608}%
. Consequently, we can double check $M=0$ by calculating the renormalized
action explicitly. From \cite{Kim1608}, we read the counterterms%
\begin{equation}
S_{\mathrm{ct}}=\int d^{3}x\sqrt{-\gamma }\left( -4+R+\frac{1}{3}\phi
n^{u}\partial _{u}\phi -\frac{1}{6}\phi ^{2}\right) .
\end{equation}%
Note that we have selected the grand canonical ensemble and the mixed
boundary condition of the dilaton has been imposed to be consistent with the
thermodynamic first law. Here $n^{u}$ is the radial component of the outward
unit vector normal to the boundary.

In order to use the formalism in the previous section, we need to change the
coordinate $u\rightarrow r$, defined by%
\begin{equation}
r=\sqrt{g_{uu}}=\left( 1+\frac{Q}{u}\right) ^{3/4}u.  \label{ur}
\end{equation}%
The exact solution is lengthy. To be explicit, we expand Eq. (\ref{ur}) at
small $Q$, which has the approximate solution%
\begin{equation}
u=\frac{1}{4}(-3Q+\sqrt{3Q^{2}+16r^{2}}).  \label{ur2}
\end{equation}%
Then we read off%
\begin{eqnarray}
L_{\mathrm{matter}}^{(1)} &=&\frac{1}{3}\left[ \phi n^{r}\partial _{r}\phi -%
\frac{1}{2}\phi ^{2}\right] , \\
L_{\mathrm{matter}}^{(2)} &=&\frac{1}{12}\left[ \frac{1}{2}\sqrt{g_{tt}}%
r^{2}\phi ^{2}-\frac{\sqrt{g_{tt}}}{\sqrt{g_{rr}}}r^{2}\phi \phi ^{\prime }%
\right] \bar{h}h, \\
L_{\mathrm{matter}}^{(3)} &=&\frac{1}{16}Q^{3}\bar{h}h.
\end{eqnarray}%
As expected, one can find%
\begin{equation}
L_{\mathrm{matter}}^{(3)}\left( \omega \right) -L_{\mathrm{matter}%
}^{(3)}\left( 0\right) =\omega ^{2}M=0.\
\end{equation}%
Also, we have checked that $M$ is still zero for large $Q$.

Then we can carry out the numerical experiments. We set the horizon radius $%
u_{+}=1$ for convenience. The performance of the DNN for the EMD black hole
is similar to that for the RN black holes.

\section{Training scheme and report}

For all numerical experiments in this paper, we implement the same training
scheme. We train the network in two stages. First, the initial weights are
randomly selected from $\left( 0,2\right) $. The loss function is given by
the sum of Eq.\ (12) and Eq. (13) of the main text. We will adopt the
RMSProp optimizer \cite{RMSProp}. Second, the initial weights of the DNN
will be replaced by the trained weights of the first stage. The loss
function is re-set as Eq. (12) of the main text without the regularization.
Then the network will be trained again with the optimizer Adam \cite{Adam}.
After the training of each stage, one can read the loss, extract the
weights, and calculate their error. It can be found that after the second
stage of training, the performance of the DNN is usually improved. In
particular, the loss (without regularizations) of the first stage can be
reduced by several orders of magnitude. Moreover, turning the regularization
factors in the first stage can improve the performance of the DNN in the
second stage. With this in mind, we will scan the parameter space of
regularization factors.

In two stages of training, we fix the batch size $c_{\mathrm{bs}}=512$, but
the learning rate is changed even not once and selected by experience. The
number of epoches in each stage is large enough so that the training will
not stop until the validation loss is almost not reduced. The regularization
factor $c_{3}$ is set as $15c_{1}$. We focus on turning the regularization
factors $c_{1}$ and $c_{2}$ because the performance of the DNN is more
sensitive to them than other hyperparameters. Initially, $c_{1}\ $and $c_{2}$
are limited to some suitable ranges. Then we scan the two-dimensional
parameter space. Considering the statistical fluctuation due to the
randomized initialization of the network, we train\ $5$ times for each set
of regularization factors \cite{Iten1807}. We gradually move and reduce the
ranges of parameters. We also gradually reduce the step sizes. Thus, the
scanning is concentrated around the $c_{1}$ and $c_{2}$ where the DNN
produces smaller losses. We stop the scanning when the step sizes $\Delta
c_{1}<10^{-5}$ and $\Delta c_{2}<0.01$.

We select the optimal regularization factors as the ones according to the
minimum loss. We read the minimum loss and the trained weights of the
network accordingly, and then calculate the MRE of the weights and the MRE
of the wideband data generated by the weights learned from narrowband data.
These quantities are considered as the final performance of the DNN. We
simply refer them as the \textquotedblleft minimum loss\textquotedblright ,
the \textquotedblleft learned metric\textquotedblright , the
\textquotedblleft MRE of learned metrics\textquotedblright , and the
\textquotedblleft MRE of generated data\textquotedblright , respectively.
They have been exhibited in FIG. 3 of the main text and in TABLE \ref{tab2}.
\begin{table}[th]
\caption{Training reports of various numerical experiments. We have five
target metrics and each of them is learned from three datasets: A $\protect%
\omega \in (0,1]$, B $\protect\omega \in (0.99,1]$, and C $\protect\omega %
\in (0.01,0.02]$. We list two optimal regularization factors and three\
quantities which characterize the performance of the DNN.}
\label{tab2}%
\begin{ruledtabular}
\begin{tabular}{ccccccc}
\ Target & \ Data & Optimal $c_{1}$ & Optimal $c_{2}$ & \ Minimum loss & MRE of learned metrics & MRE of generated data
\\ \hline
\ Schwarzschild & A & $1.20\times 10^{-3}$ & $2.02$ & $1\times 10^{-13}$ & $1\times 10^{-3}$ & / \\
\ Schwarzschild & B & $1.4\times 10^{-4}$ & $1.52$ & $6\times 10^{-13}$ & $2\times 10^{-3}$ & $5\times 10^{-6}$ \\
\ Schwarzschild & C & $3.4\times 10^{-4}$ & $1.49$ & $1\times 10^{-14}$ & $5\times 10^{-3}$ & $4\times 10^{-4}$ \\
\ RN q=1 & A & $9\times 10^{-5}$ & $2.00$ & $1\times 10^{-13}$ & $1\times 10^{-3}$ & / \\
\ RN q=1 & B & $2.6\times 10^{-4}$ & $1.30$ & $7\times 10^{-13}$ & $4\times 10^{-3}$ & $3\times 10^{-5}$ \\
\ RN q=1 & C & $1.1\times 10^{-4}$ & $0.90$ & $1\times 10^{-14}$ & $3\times 10^{-2}$ & $2\times 10^{-3}$ \\
\ RN q=2 & A & $7\times 10^{-5}$ & $1.25$ & $1\times 10^{-12}$ & $7\times 10^{-4}$ & / \\
\ RN q=2 & B & $6\times 10^{-5}$ & $1.15$ & $2\times 10^{-11}$ & $1\times 10^{-2}$ & $5\times 10^{-5}$ \\
\ RN q=2 & C & $4.2\times 10^{-4}$ & $0.50$ & $2\times 10^{-14}$ & $4\times 10^{-2}$ & $7\times 10^{-3}$ \\
\ EMD Q=1 & A & $8.5\times 10^{-4}$ & $1.85$ & $2\times 10^{-13}$ & $1\times 10^{-3}$ & / \\
\ EMD Q=1 & B & $2.2\times 10^{-4}$ & $1.20$ & $1\times 10^{-12}$ & $6\times 10^{-3}$ & $5\times 10^{-5}$ \\
\ EMD Q=1 & C & $2.6\times 10^{-4}$ & $1.00$ & $1\times 10^{-14}$ & $1\times 10^{-2}$ & $1\times 10^{-3}$ \\
\ EMD Q=5 & A & $2.5\times 10^{-4}$ & $1.16$ & $6\times 10^{-13}$ & $6\times 10^{-4}$ & / \\
\ EMD Q=5 & B & $1.1\times 10^{-4}$ & $0.85$ & $2\times 10^{-12}$ & $4\times 10^{-3}$ & $2\times 10^{-5}$ \\
\ EMD Q=5 & C & $3.0\times 10^{-4}$ & $0.40$ & $2\times 10^{-14}$ & $4\times 10^{-2}$ & $7\times 10^{-3}$ \\
\end{tabular}%

\end{ruledtabular}
\end{table}
\end{document}